# A Modified Dynamic Time Warping (MDTW) Approach and Innovative Average Non-Self Match Distance (ANSD) Method for Anomaly Detection in ECG Recordings

Xinxin Yao [1], Hua-Liang Wei [*][1,2]

[1] Department of Automatic Control and Systems Engineering, University of Sheffield, Sheffield, S1 3JD, UK

[2] INSIGNEO Institute for in silico Medicine, University of Sheffield, Sheffield, S1 3JD, UK

**Abstract:** ECGs objectively reflects the working conditions of the hearts as these signals contain vast physiological and pathological information. In this work, in order to improve the efficiency and accuracy of "best so far" time series analysis-based ECG anomaly detection methods, a novel method, comprising a modified dynamic time warping (MDTW) and an innovative average non-self match distance (ANSD) measure, is proposed for ECG anomaly detection. To evaluate the performance of the proposed method, the proposed method is applied to real ECG data selected from the MIT-BIH heartbeat database. To provide a reference for comparison, two existing anomaly detection methods, namely, brute force discord discovery (BFDD) and adaptive window discord discovery (AWDD), are also applied to the same data. The experimental results show that our proposed method outperforms BFDD and AWD.

## 1. Introduction

Cardiovascular diseases (CVDs) are the number 1 cause of death globally. According to the statistics of World Health Organization in 2016, there were almost 17.5 million people died from CVDs, representing 31% of all global deaths [1]. In recent decades, diagnosis and prevention of CVDs have become the leading concern in clinical medicine. As the most commonly used biological signals in medical field in the last few decades, ECGs are easy to collect via modern medical facilities. In clinical application, ECGs are used to determine the cardiac structure and function of patients. For CVDs, such as arrhythmia, myocardial and ischemia, they occur over a certain period with the heart of a cardiac patient does not work normally, in other words, corresponding ECG is an anomalous segment.

Anomalies are patterns in data that do not conform to a well-defined notion of normal behaviour [3]. In daily life, anomalous signals appear in various fields because of different reasons, for instance, bank card fraud in economic field [7, 8] and cyber-intrusion in networking field [4, 5]. Similarly, ECG of cardiac patient is anomaly at some periods. For example, ECG collected during abnormal work period of cardiac patient is different from normal period series [19]. Figure 1 gives a visual intuition of an anomalous segment which is highlighted by an ellipse.

Anomalies detection in ECG aims to detect the unexpected patterns from large volumes of non-anomalous data and such unexpected patterns are defined as anomalies. Nowadays, with the increasing aging population and the development of medical infrastructures, there are increasingly huge numbers of cardiac patients going to hospital for examination every day. Manual analysis (e.g. analysis by hand) of such mega data will consume huge amount of manpower and resources. Moreover, errors may occur in the detection result due to high-intensity handwork. Recently, applying machine learning and data analysis techniques to anomaly detection of ECG signals has become increasingly important in clinical medicine [15, 16, 20]. In [12], an anomaly detection method was introduced to find the most unusual time series subsequence. Later, in [14, 22, 23], a series of related works were proposed to improve the performance of the method in [12]. However, because these extended methods use the normalized Euclidean distance and non-self-match distance to find anomaly or anomalies, they cannot work well when there are two or more similar anomalies.

*Corresponding Author:* w.hualiang@sheffield.ac.uk

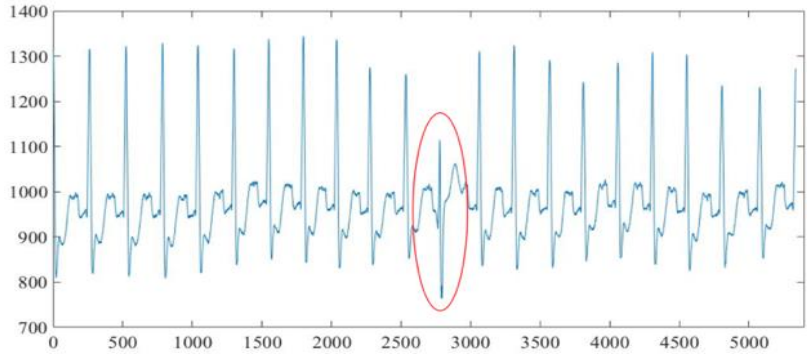

Figure 1. Anomaly detection and labelled in an ECG (collected from MIT-BIH database).

In this study, in order to correctly detect all the anomalous segments via time series mining techniques, a modified dynamic time warping (MDTW) is used to calculate the distance between candidates, and a new notion, called the average non-self match distance (ANSD), is used to detect all the anomalous segments from raw data. To demonstrate the performance of the proposed ECG anomalies detection method, both the proposed method and previous published methods are applied to 30 ECGs from MIT-BIH arrhythmia database [17]. The experimental results show that the proposed method obviously outperforms previous published methods, which is evidenced by the following anomaly detection accuracy: for non-anomalous ECGs and ECGs only contain 1 anomaly, all the three methods (BFDD, AWDD and the proposed method) work extremely well; for ECGs with 2 or more anomalies and these anomalies significantly different from each other, both AWDD and the proposed method work well while the detection accuracy of BFDD is 40%, for ECGs with more than one anomalies and these anomalies are similar to each other, BFDD and AWDD do not work (the accuracy is zero), whereas the accuracy of the proposed method is 100%.

It is worth stressing that when the proposed method is applied to ECG data, it does not just simply use RQS complex but treats the ECG data of interest as a whole to reveal any anomaly through data segmentation and the analysis of the associated dynamic time warping and the average non-self match distance using the newly proposed algorithms.

The reminder of this paper is organized as follows. In Section 2, related works are described. In Section 3, the proposed time series similarity measure method and newly defined anomalies detection notion are presented. In Section 4, the proposed ECGs anomalies detection method and existing time series analysis based approaches are applied to a series of ECGs, and some comparative analysis results are reported. Finally, this work is briefly summarized in Section 5.

## 2. Related Work

In the past few decades, anomalies detection has been used in diverse fields, such as cyber-intrusion detection [6], fraud detection [9], medical anomaly detection [10, 11], and so on. This study focuses on ECG anomaly detection via time series analysis, where ECG is regarded as a periodical time series reflecting the periodical change of heartbeat [2]. In this section, basic notion (non-self match) of time series analysis based anomaly detection is described firstly, then two popular ECG anomaly detection methods are reviewed, namely, brute force discord discovery (BFDD), and adaptive window based discord discovery (AWDD).

### 2.1 Nearest Non-Self Match

Given one time series $A$ and one of its subsequence $B$ beginning at position $P$, in general, in time series $A$, the beginning points of the best matches to $B$ (apart from itself) should be $P \pm 1$ or $P \pm 2$. Therefore, excluding unnecessary matches is an important step prior to detecting anomalies, otherwise, distance between candidate subsequence and its corresponding best match subsequence will lower than a pre-obtained threshold and it is impossible to find anomalies. In [12], one matching notion, called as *Non-Self Match*, was introduced to remove trivial matches in the process of anomaly detection.

Given one time series $T$ and its two subsequences $T1$ and $T2$, the beginning points of $T1$ and $T2$ are $P$ and $Q$, and the length values of both $T1$ and $T2$ are set as $n$, $T2$ can be defined as a non-self match to $T1$ if the position distance greater or equal to $n$ ($|P - Q| \geq n$). As an example, this is shown as below:

$$T = \fbox{$a\ b\ c$}\ a\ b\ c\ a\ b\ c\ a\ b\ c\ a\ b\ c\ \fbox{$a\ b\ c$}\ a\ b\ c\ a\ b\ c\ a\ b\ c$$

Note that $T1$ and $T2$ are labelled by dot box and solid box respectively, the length values of them are set as 3, the beginning point of $T1$ is 1 and the beginning point of $T2$ is 19. In this case, $T2$ can be defined as a non-self match to $T1$ because $|19 - 1| \geq 3$. On the contrary, another example shown in Figure 3 describes that $T2$ cannot be defined as a non-self match to $T1$.

Below is another case, where the settings are the same as in the above illustration, the only difference is that the beginning points of $T1$ and $T2$ are 17 and 19 respectively. It can be noticed that $T1$ and $T2$ are partly fold together. For the position distance, it is $|19 - 17| = 2$ and hence lower than the length of sliding window as shown as follows:

$$T = a\ b\ c\ a\ b\ c\ a\ b\ c\ a\ b\ c\ a\ b\ c\ a\ \fbox{$b\ c\ \fbox{$a$}\ b\ c$}\ a\ b\ c\ a\ b\ c\ a\ b\ c$$

Non-self match subsequences of one segment cannot be directly used to define whether the corresponding candidate is anomalous or not. For previous published anomaly detection methods, nearest non-self match of the candidate is used to detect the anomalous segments. Given one time series $T$ and one candidate $A$, the distances between $A$ and all its non-self matches are calculated and recorded as $D = [d1, d2, ..., dn]$, the nearest non-self match distance is the minimum value in $D$ and the corresponding segment is the nearest non-self match of $A$.

## 2.2 Brute Force Discord Discovery

Brute force discord discovery (BFDD) algorithm was initially proposed in [12, 13] and the advantage of this algorithm is that it is easy to understand and implemented. Based on BFDD, the implemented procedure of time series anomaly detection is as follows: 1) the first segment to be tested is the subsequence whose length is equal to that of sliding window and its first point is the same with that of time series, as shown in Figure 2(a) and Figure 2(b); 2) after the definition of testing subsequence, sliding down the defined window on sample at a time and calculating the distance between the testing subsequence and the subsequence in the sliding window, this process can help us to find the nearest non-self match of the testing subsequence, as shown in Figure 2(c); 3) once the first nearest non-self match distance and it corresponding subsequence are recorded, the first point of the testing subsequence will move from the first point of the time series to the second point, meanwhile, the length of testing subsequence is still equal to that of sliding window, as shown in Figure 2(d); 4) for every testing subsequence with length m, and an original time series with length n, in order to find the nearest non-self match, it needs to calculate at least $(n - 3m + 1)$ times, and sometimes $(n - 2m + 1)$ times, as shown in Figure 2(e). 5) according to the obtained nearest non-self match distance values, the anomalous segment can be detected via the comparison between the recorded values and a pre-obtained threshold, which is calculated by applying BFDD to training time series. Figure 2 illustrates the mechanism of BFDD.

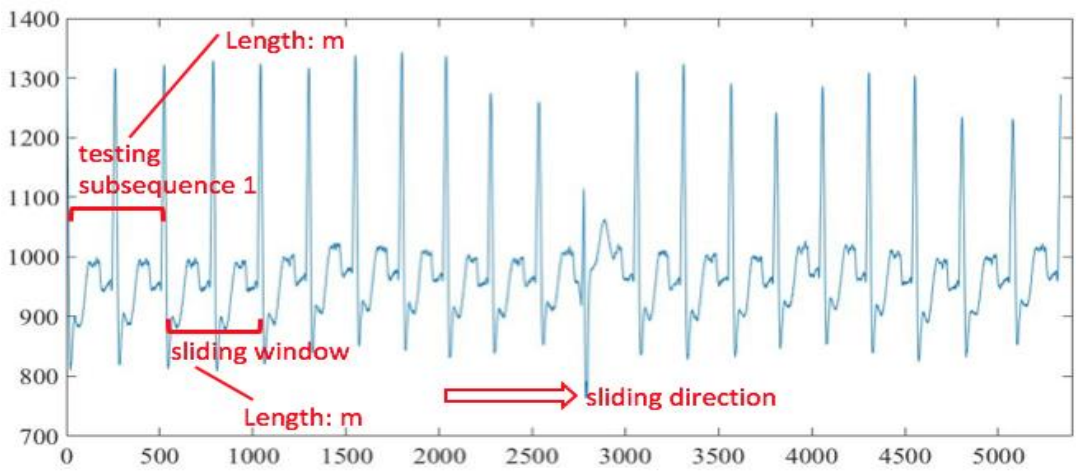

(a)

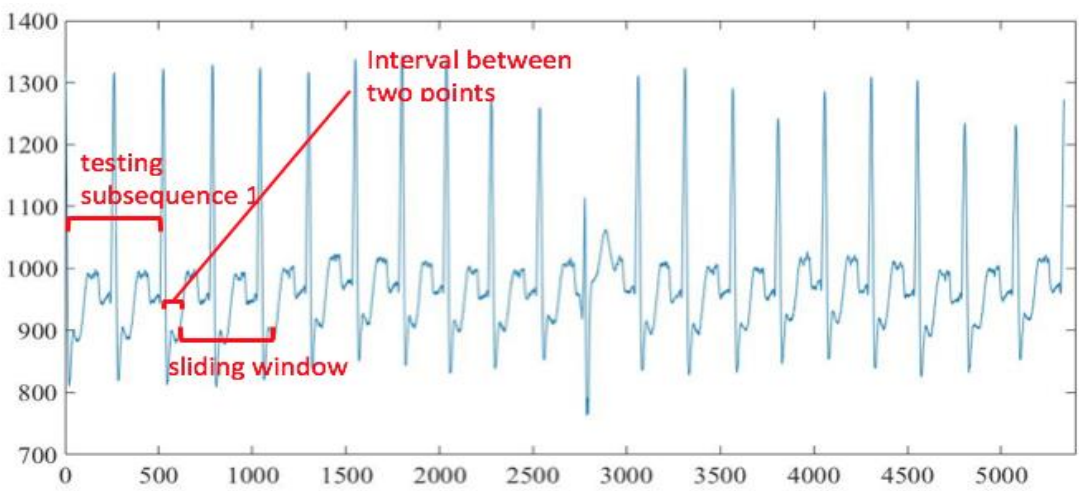

(b)

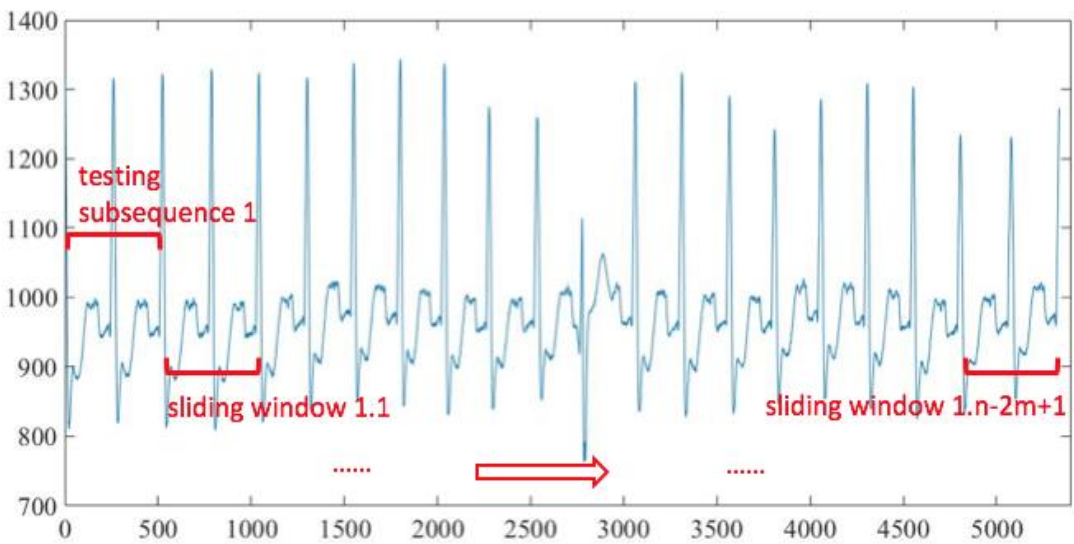

(c)

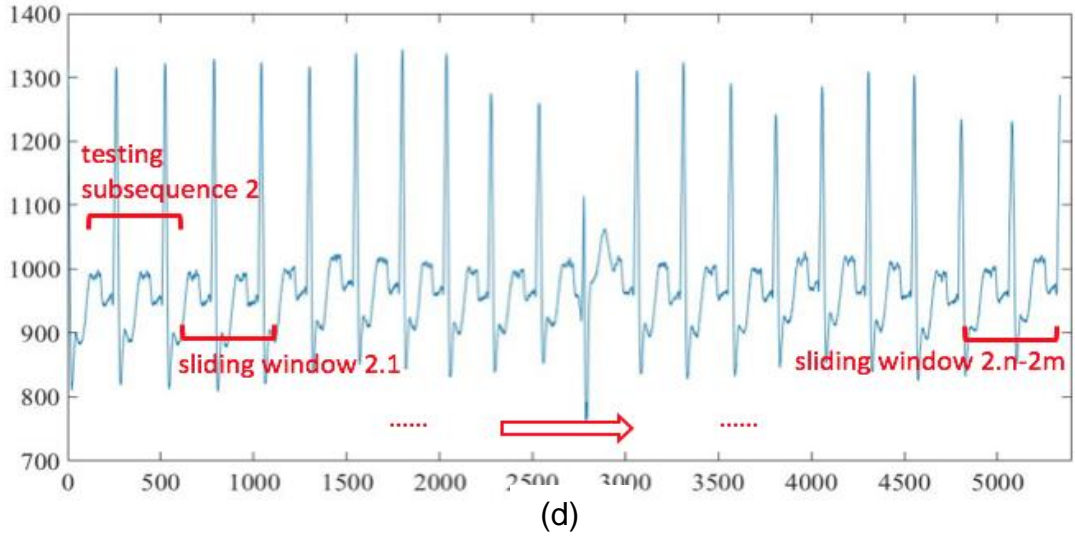

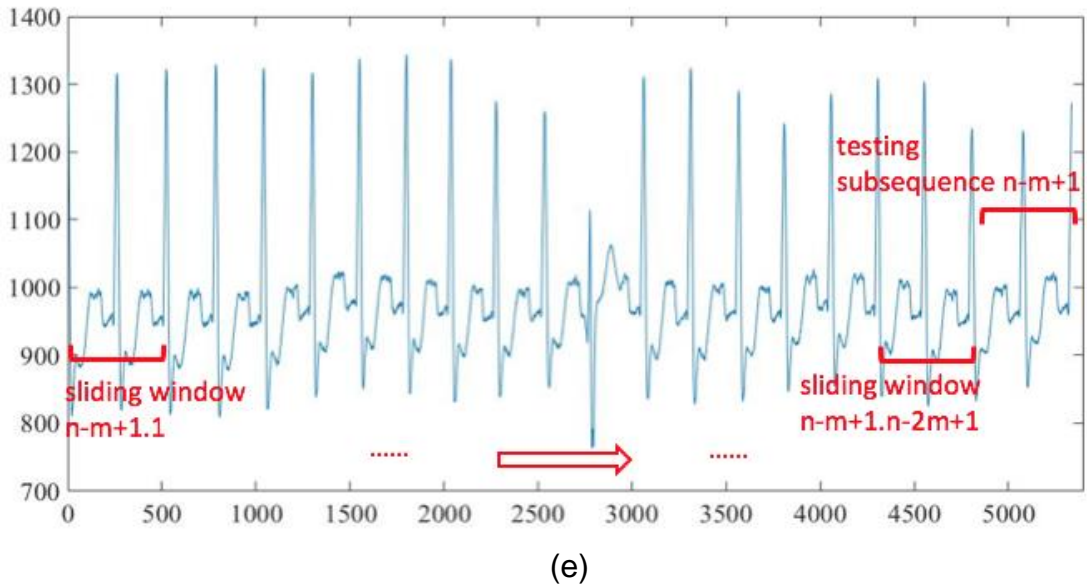

Figure 2. Mechanism of BFDD-based anomaly detection. (a) defining the first testing subsequence and sliding window; (b) sliding down the window on sample at a time; (c) every testing subsequence has at least $n - 3m + 1$ non-self match distances; (d) moving 1 unit backward of the testing subsequence; (e) the testing subsequence keeps moving to the end of the time series.

The pseudo-code and procedure of BFDD is described in Algorithm 1.

**Algorithm 1 Brute Force Discord Discovery**

**Requirements:** A time series: T

           The length of time series: n

           The length of sliding window: m

best_so_far_dist ← 0

**for** i = 1 to n − m + 1 **do**

        nearest_neighbour_dist = infinity

   **for** j = 1 to n − m + 1 **do**

      **if** |i − j| ≥ n **do**

      **if** Dist(Tp, … , T(p + n − 1), Tq, … , T(q + n − 1)) < nearest_neighbour_dist **do**

```
            nearest_neighbour_dist = Dist(Tp, ... , T(p + n − 1), Tq, ... , T(q + n − 1))
        end if
    end if
  end for
  if nearest_neighbour_dist > best_so_far_dist do
      best_so_far_dist = nearest_neighbour_dist
      best_so_far_loc = p
  end if
end for
```

From the above pseudo-code, it can be noticed that this method is achieved with nested calculations, where the outer calculation considers each possible candidate subsequence, and the inner calculation performs a linear scan to identify the nearest non-self match of corresponding candidate [28].

As an example, the length of sliding window is 300, BFDD is applied to the ECG signal shown in Figure 1, the associated nearest neighbor distances are shown in Figure 3.

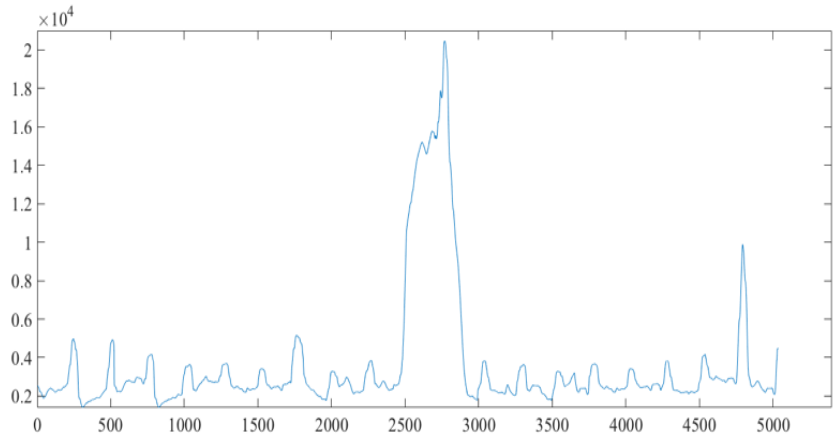

Figure 3. Nearest Neighbor Distances

Once the nearest non-self match distances are computed, all these distances will be compared with a pre-obtained threshold (computed through applying BFDD to training dataset) and the corresponding segments are going to be identified. But it is worth mentioning the low computational efficiency of BFDD-based anomaly detection. As it can be observed that there is a nested calculation during the whole process, every calculation has to compute the distances between testing subsequence and its non-self subsequence at least $(n − 3m + 1)$ times ($n$ is the length of time series and $m$ is the length of sliding window), this will take a huge amount of time for even moderately large datasets. Assuming the length values of sliding window and testing subsequence are set as 300, as the ECG signal shown in Figure 1 is a 15 seconds record and contains 5400 data points, the whole calculation process has to take over 20 million times. For a normal computer, it has to take at least 50 seconds to finish the calculation. Although BFDD-based time series anomaly detection is time consuming, there is one advantage need to be mentioned, that is universality, which means this method can be used to detect anomalies for various kinds of time series.

## 2.3 Adaptive Window Discord Discovery

For some special types of time series, the application of universal anomaly detection methods may complicate the operation process and distort the calculation accuracy. In terms of ECG data, in order to overcome the heavy computational load involved in BFDD-based anomaly detection, adaptive window discord discovery (AWDD) was proposed in [14]. AWDD separates ECG into a number of segments based on the peak points, then measures the distances between each segment and determines which subsequence can be treated as anomaly. It should be noted that the segments separated by peak points do not have same length while the distance measure in AWDD is Euclidean distance, which requires the length values of two candidates should be the same. To solve this problem, if the length values of two candidates are different, the longer one is compressed firstly so that its length is equal to that of the shorter one. In comparison with BFDD, the calculation time is significantly reduced through the application of AWDD without losing detection accuracy. The whole process of AWDD-based anomaly detection is described by Algorithm 2.

**Algorithm 2. Adaptive Window Discord Discovery**

**Requirements:** A time series: T

                    Location of peak points: P

$n \leftarrow$ length of input time series

$m \leftarrow$ number of peak points

**for** $i = 1 : n - 1$ **do**

     $outlength = P(i + 1) - P(i);$

     **for** $j = 1 : n - 1$ **do**

         $innerlength = P(j + 1) - P(j);$

         **if** $outlength > innerlength$ **do**

            $B = imresize(A(P(i) : P(i + 1)), [1, innerlength]);$

            $C = A(P(j) : P(j + 1) - 1);$

         **else**

            $B = imresize(A(P(j) : P(j + 1)), [1, outlength]);$

            $C = A(P(i) : P(i + 1) - 1);$

         **end if**

         $ddd(j) = dist(B, C)$

     **end for**

     $nearest\_neighbour\_dist(i) = min(ddd);$

     **if** $nearest\_neighbour\_dist (i) > threshold$ **do**

       $best\_so\_far\_loc = i$

     **end if**

**end for**

_______________________________________________

The inputs of Algorithm 2 include one ECG time series and the locations of peak points. The output of this Algorithm is the location or locations of anomaly or anomalies.

AWDD compresses the longer subsequence so that its length is equal to that of the shorter one, this enables the use of Euclidean distance to measure the distance between two candidates. As an illustration, Figure 4(a) provides a simple example of two time-series with different length values. In Figure 4(b), the longer time series is compressed.

AWDD is also applied to the ECG data in Figure 1. As there are 21 normal peak points, prior to distance measure, the ECG was separated into 20 segments. Then every segment is regarded as one testing subsequence and the corresponding nearest non-self match distances of these 20 segments are recorded, as shown in Figure 5.

As the result of BFDD-based anomaly detection shown in Figure 5 that nearest non-self match distances are anomalous between 2500 to 3000, the location of abnormal point of nearest non-self match distances in Figure 5 is 11 and as the interval between two normal points is almost 260, it is obvious that AWDD-based nearest non-self match distances reserve most information of BFDD-based nearest non-self match distances. For the improvement of AWDD, the distances are calculated between every segment instead of sliding down the testing subsequence on one sample time, the whole procedure of AWDD-based anomaly detection is only $(20-1)^2$ times, for a normal computer, it only takes about 0.36 seconds.

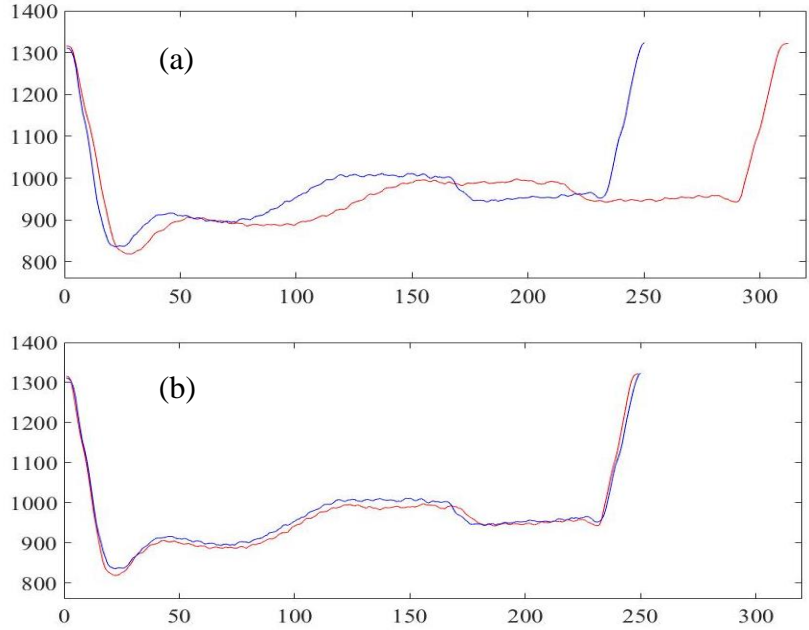

Figure 4. Two time series. (a) before compression; (b) after compression.

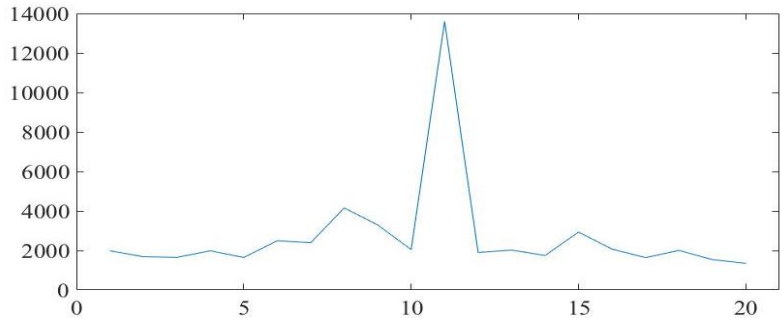

Figure 5. Nearest neighbour distance based on AWDD.

## 3. Proposed Method

If there is only one disordered segment or several significantly different disordered segments in an ECG signal, both BFDD and AEDD can correctly detect the anomalous segment(s) while AWDD outperforms BFDD in term of computational efficiency, but both methods have two common drawbacks: i) Euclidean distance measure method may influence the accuracy of anomaly detection if timeline drift exits during the process of calculating the non-self match distances; ii) they cannot correctly detect the anomalies when there are two or more anomalous segments and the distance between anomalies lower than the threshold. In this section, in order to improve the accuracy and efficiency of ECG anomalies detection, one modified distance measure method and one new notion are applied to ECG anomaly detection, as introduced as follows: 1) modified DTW is presented to improve the accuracy of time series distance measure; 2) average non-self match distance is proposed to replace nearest non-self match distance; 3) the analysis procedure using the proposed method for ECG anomaly detection is described.

## 3.1 Distance Measure

Unlike traditional DTW-based distance measure (which is detailed in [18, 21]), where DTW is used to directly calculate the distances between corresponding points and sum them as final distance. In this paper, the distance between two candidates is defined according to the DTW distance and the optimal align. Given two time series A and B, the distance between them is defined as follows:

$$\text{Dist}(A, B) = d + \left(\frac{l - la}{la}\right) \times \big(\text{sum}(Anew) - \text{sun}(A)\big)$$
$$+ (\frac{l-lb}{lb}) \times (\text{sum}(Bnew) - \text{sum}(B)) \quad (1)$$

where $\text{Dist}(A, B)$ represents the distance between A and B, d is the DTW distance between A and B, l is the length of the optimal align path; Anew and Bnew are two new time series segments which are constructed according to A, B and the optimal align path; la and lb are the length values of Anew and Bnew; the function $\text{sum}(\dots)$ returns the sum of elements of the input segment. The whole process of this distance measure method is summarized by Algorithm 3.

**Algorithm 3 Distance Calculation**

**Requirements**: Two time series A and B

la ← length of A
lb ← length of B
distance ← DTWdistance $(A, B)$
optimal path ← DTWdrift $(A, B)$
wa ← first column of optimal path
wb ← second column of optimal path
l ← length of optimal path
**for** $i = 1$ to l **do**
    Anew(i) = A(wa(i))
    Bnew(i) = B(wb(i))
**end for**
 final distance = distance + $((l - la)/la) \times (\text{sum}(Anew)$

      $-\text{sum}(A)) + ((l - lb)/lb) \times (\text{sum}(Bnew) - \text{sun}(B))$

_______________________

The inputs of this algorithm are two time series and the output is the distance between them.

To demonstrate the performance of the proposed method for time series similarity measure, two time series A and B are shown in Figure 6. The proposed method, together with traditional dynamic time warping and Euclidean distance are applied to calculate the distance between A and B.

| A | 1 | 2 | 3 | 4 | 5 | 6 | 6 | 7 | 8 | 9 | 10 | 11 | 12 | 13 | 13 |
|---|---|---|---|---|---|---|---|---|---|---|----|----|----|----|----|
| B | 1 | 2 | 3 | 4 | 5 | 6 | 7 | 8 | 9 | 9 | 9 | 10 | 11 | 12 | 13 |

Figure 6. Template Time Series

The calculation matching image of Euclidean distance is shown in Figure 7 and the distance between A and B is 7. However, we can find that the subsequence from 7th point to 10th point in time series A is similar to the subsequence from 6th point to 9th point in time series B, but Euclidean distance directly calculate the distances between elements at same time point and sum them up as the distance between these two time series.

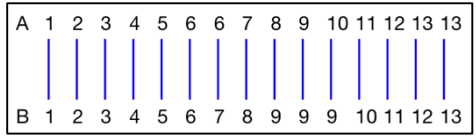

Figure 7. Matching Image of Distance Calculation based on Euclidean Distance.

The matching image of using traditional DTW to measure the distance between A and B is shown in Figure 8. It can be noticed that the timeline has been warped and the most similar elements have aligned with each other. However, the final distance between these two time series is 0 although they are not identical.

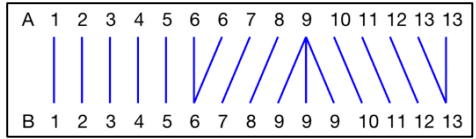

Figure 8. Matching Image of Distance Calculation based on DTW

Traditional distance measures such as Euclidean distance and DTW do not work well for the above time series A and B. That is the motivation to propose the new method (Algorithm 3) to overcome the disadvantage of traditional distance measures. In Algorithm 3, in order to eliminate the impact of the neglect of timeline drift, the distance between two candidates is defined according to three variables: DTW distance, optimal align path between two time series, and the sum of distances between the extended new points and base point (these new points exist when there is timeline drift between two candidates, and vice versa). Compare with the results obtained based on Euclidean distance and traditional DTW, the distance between A and B computed based on Algorithm 3 is 4.9333, which not only eliminate the error caused by existence of timeline drift, but also remove the error caused by neglect of timeline drift.

### 3.2 Average Non-self Match Distance

To overcome the drawback of BFDD and AWDD that they can only work well for anomaly detection when all the anomalies in time series of interest are significantly different from each other, this subsection proposes a new notion, called as average non-self match distance.

Given one time series $T$, one of its segment is $A$, non-self matches of $A$ in $T$ are stored in $A_m = [A_1, A_2, ..., A_n]$, and the distacnes between $A$ and all its non-self matches are obtained through the application of the proposed distance measure method and recorded in $D = [d_1, d_2, ..., d_n]$. In terms of anomaly detection based on BFDD and AWDD, the minimum value in $D$ is recorded as nearest non-self match and used to identify whether $A$ is anomaly or not. Different from nearest non-self match, average value of $D$ is recorded and used in our proposed method to identify whether $A$ is anomaly or not.

In order to clearly state the advantage of average non-self match distance in anomalies detection of multi-anomalous time series, a time series contains two same anomalies is defined in Figure 9, which is constructed through repeatedly using a time series (e.g. ECG signal in Figure 1).

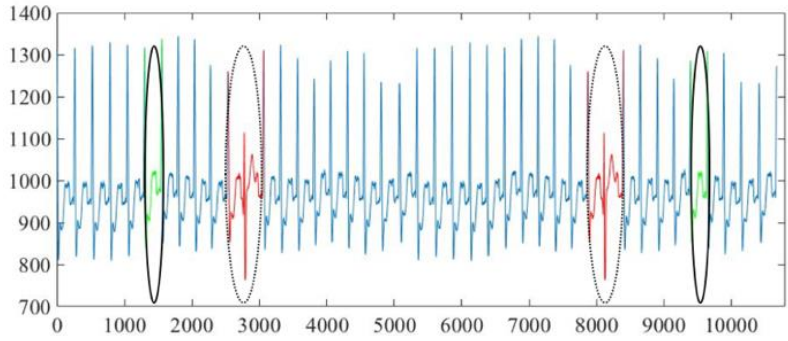

Figure 9. Two-Anomaly Time Series

Two normal segments and two anomalous segments (highlighted by solid and dot boxes in Figure 9) are extracted as testing segments, both nearest non-self match and average non-self match distance are applied to these four segments. Table 1 illustrates the values that are used to identify whether the input segments are anomalous or not.

In Table 1, the second row states the distances between the extracted segments and their corresponding nearest non-self match segments, the third row illustrates the average values of distances between extracted segments and all their corresponding non-self match segments. The values in Table 1 show that anomaly detection methods based on nearest non-self match cannot correctly detect anomalies in some special conditions, while our proposed method correctly detect all the anomalous segments.

Table 1. Values Used for Anomalies Identification

| Distance | Normal 1 | Normal 2 | Anomaly 1 | Anomaly 2 |
|---|---|---|---|---|
| Nearest Non-self Match | 0 | 0 | 0 | 0 |
| Average Non-self Match Distance | $1.9730 \times 10^5$ | $1.8508 \times 10^5$ | $6.6383 \times 10^5$ | $6.6383 \times 10^5$ |

### 3.3 Anomalies Detection in ECG Data

The analysis procedure using the proposed method for ECG anomaly detection is as follows: 1) separate the input ECG into several segments based on peak points; 2) define the anomalies using the average non-self match distance of every segment.

#### 3.3.1 Peak points Collection

It is known that ECG can be defined as periodical time series because ECG derives from regularly heart muscle beat. Therefore, peak points based ECG segmentation is applied prior to distance measure. Algorithm 4 briefly describes the procedure of peak points collection.

**Algorithm 4 Peak Points Collection**

**Requirements:** An ECG signal: T
                 A defined value: h
n ← length of input ECG signal
m ← 1
**for** i = 1 to n **do**
    **if** T(i) ≥ h **do**
        location(m) = i
        m ← m + 1
    **end if**
**end for**

The inputs of Algorithm 4 include one ECG signal and one threshold, this threshold is obtained through training the available ECG data. The output is a vector contains locations of peak points.

As an example, Algorithm 4 is applied to the ECG signal shown in Figure 1, and the detected peak points are highlighted by red dots and shown in Figure 10.

### 3.3.2 Anomaly Detection

As mentioned in Introduction, anomalies are patterns in data that do not conform to a well-defined notion of normal behaviour. Therefore, at the beginning of anomaly detection, a criterion has to be defined and it can be obtained through applying the proposed anomaly detection method to available training data. For example, average non-self match distances of all normal segments are stored in $D_n$, average non-self match distances of all anomalous segments are stored in $D_a$, the threshold is always defined as the average value of the maximum value in $D_n$ and the minimum value in $D_a$.

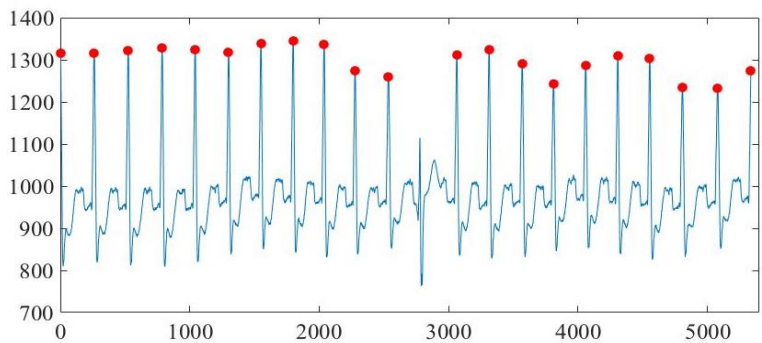

Figure 10. Peak Points Collection

Once the threshold is obtained, the average non-self match distances of all the segments in testing signal have to be computed and compared with the threshold, if the value greater than the threshold, the corresponding segment is defined as anomalous, on the contrary, if the value lower than the threshold, the corresponding segment is defined as normal. Figure 11 shows the average non-self distances of all the segments in ECG signal in Figure1, as the threshold is $1.8 \times 10^4$ (which is obtained through applying the proposed method to 10 training ECG signals), the average non-self match distance of the 11st segment is greater than the threshold, hence the 11st segment is anomaly and the others are normal segments.

It can be noticed that the average non-self match value of the anomalous segment in Figure 11 is significantly greater than the values of normal segments, and the corresponding segment can be defined as anomalous directly without the comparison with the threshold. But it should be noted this is a special condition, in some common cases, comparing with threshold is the best way to correctly detect the anomalous segments.

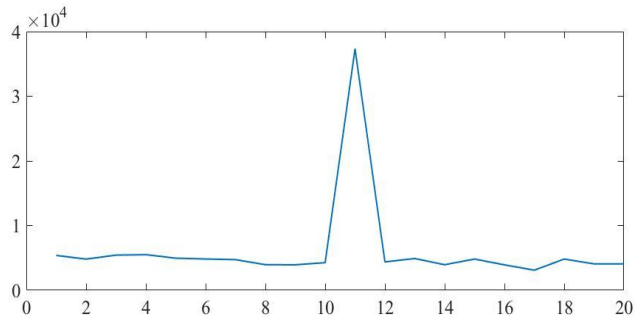

Figure 11. Average Non-self Match Distances of Time Series in Figure 1

## 4. Case Study

The resource of ECGs (30 ECGs in total) included in the MIT-BIH database is a set of over 4000 long-term Holter recordings that were obtained by Beth Israel Hospital Arrhythmia Laboratory from 1975 to 1979. Approximately 60% of the recordings were obtained from inpatients [34]. Because of these, testing performance of ECG anomaly detection algorithms based on this database has strong conviction. The experimental database used in this study contains 30 ECGs, 10 of them are used as training database to obtain the threshold and the rest 20 ECGs are used as testing database.

Table 2. ECG Excerpts from MIT-BIH Record 109

| No. ECG Datasets for Training | Start-end points | Anomaly Location | Anomaly Identification |
|---|---|---|---|
| 1 | 140s-180s | NA | NO |
| 2 | 440s-480s | NA | NO |
| 3 | 560s-600s | NA | NO |
| 4 | 700s-740s | NA | NO |
| 5 | 740s-780s | NA | NO |
| 6 | 20s-60s | 6758 | YES |
| 7 | 80s-120s | 4474 | YES |
| 8 | 200s-240s | 12890 | YES |
| 9 | 260s-300s | 5690 | YES |
| 10 | 520s-560s | 3205 | YES |
| **No. ECG Datasets for Testing** | **Start-end points** | **Anomaly Location** | **Anomaly Identification** |
| 11 | 860s-900s | NA | NO |
| 12 | 940s-980s | NA | NO |
| 13 | 980s-1020s | NA | NO |
| 14 | 1180s-1220s | NA | NO |
| 15 | 1220s-1260s | NA | NO |
| 16 | 620s-660s | 11630 | YES |
| 17 | 660s-700s | 7928 | YES |
| 18 | 820s-860s | 9410 | YES |
| 19 | 1300s-1340s | 6100 | YES |
| 20 | 1400s-1440s | 7170 | YES |
| 21 | 900s-940s | 6270,11590 | YES |
| 22 | 1060s-1100s | 10320,12920 | YES |
| 23 | 1100s-1140s | 2412,11970 | YES |
| 24 | 1140s-1180s | 966,9957 | YES |
| 25 | 500s-540s | 3655,10410 | YES |
| 26 | NA | 11630, 26030 | YES |
| 27 | NA | 7928, 22328 | YES |
| 28 | NA | 9410, 23810 | YES |
| 29 | NA | 6100, 20500 | YES |
| 30 | NA | 7170, 21570 | YES |

The 30 ECGs used in the study are listed in Table 2, where the 1st column is an index of the 10 training ECGs and 20 testing ECGs, the 2nd column illustrates the starting and ending time of corresponding ECG signal, the 3rd column is a location index to show where the anomaly occurs, and the 4th column is an indication of if there is an anomaly or anomalies in the corresponding ECG. Each of the first 5 training ECGs does not contain anomalous segment whereas each of the last 5 training ECGs contains one anomalous segment. In testing dataset, in order to demonstrate the ability of the proposed method for detecting multiple anomalous ECGs, the 20 test ECGs were chosen as follows: No. 11-15 contain no anomaly, No. 16-20 contain one anomaly each, No. 21-25 contain 2 significantly different anomalies, and No. 26-30 are constructed through repeating one-anomalous (No. 16-20) segment and each of them contains 2 same anomalies.

## 4.1 BFDD Based Anomaly Detection

BFDD is applied to training ECGs (No. 1-10 in Table 2) to calculate the threshold. The nearest non-self match distances of normal segments and anomalous segments are shown in Table 3.

Table 3. Threshold Calculation based on BFDD

| No. ECG Datasets | Maximum Nearest Non-self Match Distances | Anomaly |
|---|---|---|
| 1 | 5102 | NO |
| 2 | 4886 | NO |
| 3 | 9206 | NO |
| 4 | 5582 | NO |
| 5 | 6056 | NO |
| 6 | 21171 | YES |
| 7 | 22469 | YES |
| 8 | 16947 | YES |
| 9 | 9996 | YES |
| 10 | 22162 | YES |

On the basis of the obtained values, the threshold has to clearly tell whether the segment is an anomaly or not. Table 3 shows that the maximum value in the second column in relation to the normal segments is 9206, and the minimum value in the second column in relation to the anomalous segments is 9996. Therefore, the threshold is defined as the average value of 9996 and 9206, that is 9601.

With the threshold, BFDD-based anomaly detection is applied to testing ECGs. The first step is to calculate and record the nearest non-self match distance of every sliding window, and the compare the recorded values with the threshold to identify the corresponding segment. Table 4 shows the anomaly detection results of the 20 testing ECGs.

Table 4. Anomaly Detection based on BFDD

| No. ECG Datasets | Detected Location | Maximum Nearest Non-self Match Distance | Anomaly Identification | Operation Time (seconds) |
|---|---|---|---|---|
| 11 | NA | 6093 | NO | 461.8 |
| 12 | NA | 7299 | NO | 458.6 |
| 13 | NA | 5351 | NO | 461.9 |
| 14 | NA | 5634 | NO | 459.2 |
| 15 | NA | 3831 | NO | 456.3 |
| 16 | 11673 | 24023 | YES | 458.8 |
| 17 | 7942 | 22513 | YES | 461.8 |
| 18 | 9421 | 17365 | YES | 455.1 |
| 19 | 6106 | 23793 | YES | 459.6 |
| 20 | 7178 | 20911 | YES | 460.7 |
| 21 | 6286, 11611 | 21261, 19876 | YES | 459.1 |
| 22 | 6750, 7659 | 8201, 6239 | YES | 457.8 |
| 23 | 4679, 7730 | 7299, 9206 | YES | 461.9 |
| 24 | 971, 9960 | 18481, 17365 | YES | 457.8 |
| 25 | 6089, 9409 | 6285, 6250 | YES | 460.7 |
| 26 | NA | 0 | NO | 1719.8 |
| 27 | NA | 0 | NO | 1843.5 |
| 28 | NA | 0 | NO | 1843.8 |
| 29 | NA | 0 | NO | 1855.1 |
| 30 | NA | 0 | NO | 1863.9 |

In Table 4, the 2nd column shows the location or locations of the detected anomaly or anomalies, the 3rd column illustrates the maximum nearest non-self match distance or the nearest non-self match distances that greater than the threshold, the 4th column describes the results of anomaly detection and the last column lists the calculation time used by BFDD. It can be seen that BFDD can correctly define that the ECG is normal or anomalous when there is no anomaly or only one anomaly in one ECG. When there are two significantly different anomalous segments in testing ECG, BFDD can also detect the presences of anomalies, but the accuracy is only 40%. What is worse is when there are two same or similar anomalies, BFDD cannot detect any of them. In terms of computation complexity, BFDD-based anomaly detection is not acceptable, as the length of testing ECG only contains the records in 40 seconds, the calculation time is over 450 seconds.

## 4.2 AWDD Based Anomaly Detection

AWDD-based anomaly detection is applied to the same ECG data. Based on Algorithm 2, the first step is the same with that of BFDD-based anomaly detection, which is to define the threshold through applying AWDD to training dataset. In this subsection, the threshold for ECG anomaly detection is 8325. The results of applying AWDD to training ECGs are shown in Table 5.

Table 5. Threshold Calculation based on AWDD

| No ECG Datasets | Maximum Nearest Non-self Match Distances | Anomaly |
|---|---|---|
| 1 | 2326 | NO |
| 2 | 1960 | NO |
| 3 | 2636 | NO |
| 4 | 2133 | NO |
| 5 | 3068 | NO |
| 6 | 1433 | YES |
| 7 | 15990 | YES |
| 8 | 13583 | YES |
| 9 | 13830 | YES |
| 10 | 15164 | YES |

Once the threshold is known, AWDD is then applied to the 20 testing ECGs. Table 6 shows the results of the AWDD-based anomaly detection.

Table 6. Anomaly Detection based on AWDD

| No. ECG Datasets | Detected Location | Maximum Nearest Non-self Match Distance | Anomaly Identification | Operation Time |
|---|---|---|---|---|
| 11 | NA | 2224 | NO | 1.9752 |
| 12 | NA | 2006 | NO | 1.8634 |
| 13 | NA | 3670 | NO | 1.7463 |
| 14 | NA | 1873 | NO | 1.8055 |
| 15 | NA | 1630 | NO | 1.9323 |
| 16 | 11440 | 15650 | YES | 2.1261 |
| 17 | 7800 | 15346 | YES | 1.9347 |
| 18 | 9360 | 13206 | YES | 1.9914 |
| 19 | 5880 | 15263 | YES | 1.7558 |
| 20 | 7020 | 14696 | YES | 1.9283 |
| 21 | 6240, 11440 | 13008, 12578 | YES | 1.9469 |
| 22 | 10140, 12740 | 15788, 15678 | YES | 1.9222 |
| 23 | 2340, 11960 | 11520, 13250 | YES | 2.2854 |
| 24 | 780, 9800 | 14967, 12191 | YES | 1.8531 |
| 25 | 3640, 10400 | 8542, 10254 | YES | 1.8206 |
| 26 | NA | 0 | NO | 6.9522 |
| 27 | NA | 0 | NO | 7.0700 |
| 28 | NA | 0 | NO | 7.3153 |
| 29 | NA | 0 | NO | 7.8347 |
| 30 | NA | 0 | NO | 7.3965 |

Table 6 shows that AWDD can correctly tell the normal ECGs (No. 11-15). For testing ECGs (No. 16-20), AWDD can also correctly detect anomalous segments and identify their corresponding location. When there are two different anomalies in testing ECGs (No. 21-25), AWDD can detect the existences of all the anomalies, which is outperforms the results of BFDD-based anomaly detection, while for the rest 5 testing ECGs (No. 26-30), the results are same with BFDD-based anomaly detection, no anomaly is detected. One improvement has to be mentioned is that the whole process of anomaly detection for every ECGs only takes about 1.5 seconds. To sum up, AWDD is more trustworthy and efficient when compare with BFDD in terms of ECG anomaly detection.

## 4.3 Proposed Method Based Anomaly Detection

The whole process of the proposed method based ECG anomaly detection is similar with that of BFDD and AWDD. Specifically, the first step is to compute a threshold, which is obtained through training the available ECGs, the second step is to calculate the average non-self match distances of testing ECGs and identify the testing ECGs through comparing the distances with the threshold. The results generated by applying the proposed method to training ECGs are shown in Table 7.

Table 7. Threshold Calculation Based on Proposed Method

| No. ECG Datasets | Maximum Nearest Non-self Match Distances | Anomaly |
|---|---|---|
| 1 | 5308 | NO |
| 2 | 5531 | NO |
| 3 | 5832 | NO |
| 4 | 7012 | NO |
| 5 | 5246 | NO |
| 6 | 32031 | YES |
| 7 | 30694 | YES |
| 8 | 30120 | YES |
| 9 | 35691 | YES |
| 10 | 29098 | YES |

As shown in Table 7, a threshold can be computed to allow us to define whether the testing segment is anomaly or not. At here, the threshold is 18055. With the threshold, the new method is applied to testing ECGs. The results are shown in Table 8.

Table 8. Anomaly Detection Based on Proposed Method

| No. ECG Datasets | Detected Location | Maximum Nearest Non-self Match Distance | Anomaly Identification | Operation Time |
|---|---|---|---|---|
| 11 | NA | 4143 | NO | 3.6727 |
| 12 | NA | 7199 | NO | 3.3580 |
| 13 | NA | 4865 | NO | 3.1531 |
| 14 | NA | 4040 | NO | 3.3303 |
| 15 | NA | 5149 | NO | 3.1878 |
| 16 | 11440 | 34477 | YES | 3.3558 |
| 17 | 7800 | 33981 | YES | 3.1755 |
| 18 | 9360 | 32119 | YES | 3.3403 |
| 19 | 5880 | 36618 | YES | 3.4315 |
| 20 | 7020 | 41718 | YES | 3.3230 |
| 21 | 6240, 11440 | 27768, 26318 | YES | 3.3121 |
| 22 | 10140, 12740 | 32760, 31071 | YES | 3.2435 |
| 23 | 2340, 11960 | 38546, 34453 | YES | 3.3711 |
| 24 | 780, 9800 | 35035, 28146 | YES | 3.1587 |
| 25 | 3640, 10400 | 37826, 36183 | YES | 3.2899 |

| 26 | 10400, 26000 | 34527, 34527 | YES | 12.5846 |
| 27 | 7800, 22100 | 33916, 33916 | YES | 12.3937 |
| 28 | 9360, 23660 | 31946, 31946 | YES | 12.8550 |
| 29 | 5880, 20280 | 36730, 36730 | YES | 12.5591 |
| 30 | 7020, 21580 | 41759, 41759 | YES | 12.8025 |

From Table 8, it is clear that the new method has the ability to detect all the anomalies when there are more than 1 similar or same anomalies in an ECG. For the 5 ECGs that each of them contains 2 different anomalies and the 5 ECGs contain 1 anomalous segment, this new method can correctly detect the existence or existences of anomaly or anomalies. For the rest non-anomalous ECGs, this new method can also correctly identify that they are normal.

As shown in Table 4, Table 6 and Table 8, BFDD and AWDD cannot detect anomalies when there are two or more similar or same anomalies in one ECG, while the proposed method can correctly detect all the anomalies. For ECG contains two or more significantly different anomalies, BFDD-based anomaly detection has the accuracy of 40%, while the proposed method and AWDD has the accuracy of 100%. For ECG only contains one anomalous segment and non-anomalous ECG, these three methods work well, with an accuracy of 100% for all of them. In terms of computation complexity, BFDD anomaly has to take over 460 seconds while AWDD and the proposed method only take 1.9 seconds and 3.5 seconds respectively. To sum up, the proposed methods provides a promising improvement in terms of detecting anomalies from ECG signals. The overall performances of the three methods are briefly summarized in Table 9.

Table 9. Anomaly Detection Accuracy Comparison

|  | 2 or more anomalies similar or same with each other | 2 or more anomalies significantly different from each other | 1 anomaly ECG | Non-anomalous ECG |
| --- | --- | --- | --- | --- |
| New Method | 100% | 100% | 100% | 100% |
| BFDD [12] | 0 | 40% | 100% | 100% |
| AWDD [14] | 0 | 100% | 100% | 100% |

## 5. Conclusions

Given the fact that cardiovascular disease has been a focus in society and clinical fields for ages, we believe that the application of data mining method to ECG anomaly detection will make a great contribution to the heart disease detection. With the natures of fast calculation and high accuracy, data mining methods is helpful for patients to get fast and accurate treatment. In this paper, we proposed an ECG anomaly detection method based on a new distance measure method and a new anomaly detection notion. For the proposed distance measure method, with the purpose of eliminate the error caused by existence of timeline drift and remove the error caused by neglect of timeline drift, the distance between two candidates is calculated according to their DTW distance and the optimal align path between them. For the new anomaly detection notion, in order to correctly detect all the anomalies in one time series, the average value of non-self match distance is used to replace the minimum value of non-self match distances. Through applying the proposed method and the other two famous anomaly detection methods to 30 real ECGs, experimental results show that the proposed method is promising in terms of calculation complexity and outperforms the two compared methods with regards to the accuracy of anomalies detection.

## 6. Acknowledgements

This work was supported in part by the Natural Environment Research Council (NERC) under the Grant NE/V001787 and Grant NE/V002511, the Engineering and Physical Sciences Research Council (EPSRC) under Grant EP/I011056/1, the EPSRC Platform Grant EP/H00453X/1.